\documentstyle[prl,multicol,aps,psfig]{revtex}

\begin{document}

\title{ Coexistence of solutions in dynamical mean-field theory of the Mott
transition }
\author{Werner Krauth
\footnote{krauth@physique.ens.fr,
http://www.lps.ens.fr/$\tilde{\;}$krauth} }
\address{CNRS-Laboratoire de Physique Statistique\\
Ecole Normale Sup{\'{e}}rieure,
24, rue Lhomond, 75231 Paris Cedex 05, France}
\maketitle

\begin{abstract}
In this paper, I discuss the finite-temperature metal-insulator
transition of the paramagnetic Hubbard model within dynamical
mean-field theory. I show that coexisting solutions, the hallmark
of such a transition, can be obtained in a consistent way both from
Quantum Monte Carlo (QMC) simulations and from the Exact Diagonalization 
method.
I pay special attention to discretization errors within QMC. These
errors  explain why it is difficult to obtain the solutions by
QMC close  to the boundaries of the coexistence region.
\end{abstract}

\begin{multicols}{2}
\narrowtext
The Mott-Hubbard metal-insulator transition is one of the fundamental
problems in the field of correlated electronic systems. During the
last few years, dynamical mean-field theory (DMFT) \cite{DMFTrefs}
\cite{revue} has emerged as an appropriate  paradigm for this
transition. Within DMFT, the paramagnetic half-filled Hubbard model
on the Bethe lattice was found to undergo a first-order phase
transition at finite temperature $T= 1/\beta$.  The hallmark of
such a transition is the coexistence of two solutions for the same
 value of the electronic interaction $U$ with $U_{c_1}(\beta) < U
< U_{c_2}(\beta)$. As is usual, I will label ``insulating'' the
solution which can be continuously followed at low temperature from
large values of the interaction down to $U_{c_1}$ and ``metallic''
the one which exists from $U=0$ up to $U_{c_2}$.

Recent work \cite{schlipf} has challenged the above scenario.
Using the Quantum Monte Carlo (QMC) method, rather than the Exact
Diagonalization (ED) approach \cite{caffarel}, the authors of
\cite{schlipf}  were unable to detect coexistence in the DMFT
equations.  This is surprising, as the two algorithms were compared
extensively to each other (cf, e.g.  ref. \cite{revue}, section
VI, Appendix C). The aim of the present paper is to discuss the
conflicting numerical approaches (for related work, cf \cite{rozenberg}).
I show that coexisting metallic and insulating solutions can indeed
be obtained using both methods in a transparent and completely
consistent way. I pay special attention to the problem of discretization
errors.  It will become clear that the QMC method needs to use very
fine discretizations in order to obtain the insulating solution
close to $U_{c_1}$.  Inside the coexistence region, the problem
disappears.  All the numerical work in this paper relies on programs
which are publicly available \cite{revue}.

Given the large number of articles already published on the subject,
I will not repeat the  standard definitions for the half-filled
Hubbard model on the Bethe lattice within DMFT.  These can be found
e.g. in ref. \cite{revue}. The important parameter of the model is
the bandwidth $D$ (cf. \cite{revue}); I quote interactions and
temperatures in units of the bandwidth $D/\sqrt{2}$.

The point of departure of the present paper is ref. \cite{laloux},
where coexisting solutions of the mean-field equations at inverse
temperature $\beta  D/\sqrt{2} =100$  were found (within ED) for $U_{c_1}< U <
U_{c_2}$ with $U_{c_1}(\beta D/\sqrt{2} =100) \sim 3.3  D/\sqrt{2}$ and 
$U_{c_2}(\beta  D/\sqrt{2}=100)=3.8  D/\sqrt{2}$.

For this paper, I choose as a reference point the value  $U=3.55  D/\sqrt{2}$,
$\beta  D/\sqrt{2}=100$. At this point, both metallic and insulating solutions
can be found easily by ED \cite{footweb}. These solutions are given by 
bath Green's functions ${\cal G}_0(\tau)$ and impurity model 
Green's functions $G(\tau)$ (cf \cite{revue} for definitions).  

As we are well within the coexistence region, both solutions correspond
to the bottoms of deep basins of attraction.  I have, e.g., perturbed
both the metallic bath Green's function, ${\cal G}_0^{met}(\tau)$, as
well as  the insulating one, ${\cal G}_0^{ins}(\tau)$, and fallen
back into the corresponding solutions after a few iterations.  This
situation changes as we approach the boundary of the coexistence
region. For smaller $U$, for example, the insulating solution
becomes less and less attractive, until the basin of attraction
vanishes at $U_{c_1}$. This scenario simply corresponds to the
familiar free-energy landscape at a first-order transition.

Both the ED and the QMC discretize some component of  the DMFT
equations. In ED, the discretization parameter is the number $n_s$
of {\em sites} of the quantum impurity model, while in QMC the
number $L$ of {\em slices}  appears, with $\Delta \tau = \beta/L$
(cf. \cite{revue}).  I have computed the insulating bath Green's
function ${\cal G}_0^{ins}(\tau)$ at the reference point for $n_s=5,6,7$.
Each of the solutions (for fixed $n_s$) is fully converged, in
addition, the convergence with the number of sites $n_s$ is excellent.
I stress in passing that the sites in the ED algorithm are chosen
in an optimal way, very similar to what is done in Gaussian
integration \cite{recipes}. Exponential convergence in $n_s$ has
been reported \cite{revue}. Given that ${\cal G}_0^{ins}(\tau,n_s=7)$
differs by less than $0.005$  from ${\cal G}_0(\tau,n_s=6)$, I am
led to the proposition that both ${\cal G}_0^{ins}(\tau,n_s=7)$
and ${\cal G}_0^{met}(\tau,n_s=7)$ are essentially exact.  This
assertion would have to be refuted for at least one of the solutions
(in fact, for the insulating one) if we were to agree with the
authors of ref.  \cite{schlipf}.

The following reasoning will lead us to a very
interesting result:  consider a {\em single} iteration  of the
self-consistency loop  
\begin{equation} 
{\cal G}_0^{ins,met} \longrightarrow 
G \longrightarrow {\cal G}_0^{\mbox{new}} \label{loop}
\end{equation} 
If, contrary to our assertion, 
${\cal G}_0^{ins}(\tau)$ or 
${\cal G}_0^{met}(\tau)$  
were not self-consistent, we should be able to
detect differences, say,  between ${\cal G}_0^{ins}$ and ${\cal
G}_0^{\mbox{new}}$ - even by other methods than ED, for example by
QMC simulation.

In this sense, I have discretized both ${\cal G}_0^{ins}$
and ${\cal G}_0^{met}$ and used them as inputs for single self-consistency
loops (as in eq.  (\ref{loop})) of the QMC algorithm for different
values of $\Delta \tau$.  From the discretized Green's functions
$G^{\Delta \tau}(\tau)$, we then compute ${\cal  G}_0^{\mbox{new}}(\tau)$
by inverse Fourier transformation.  Results for the metallic solution
are shown in Fig.  1. We see that the finite-$\Delta \tau$ effects
are quite small. Clearly, ${\cal  G}_0^{\mbox{new}}(\tau,\Delta \tau)
\rightarrow {\cal G}_0^{met}(\tau) $ as $\Delta \tau \rightarrow 0$.   
The metallic solution has not been contested.

\begin{figure}
\centerline{ \psfig{figure=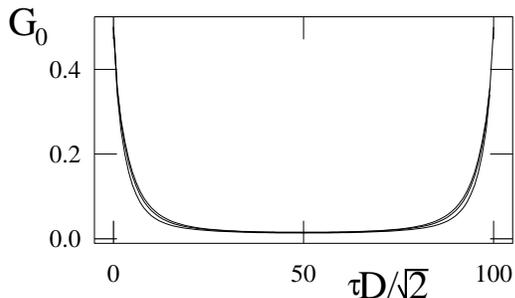,height=4cm} }
\caption{ 
${\cal  G}_0^{\mbox{new}}(\tau,\Delta \tau)$ of the metallic solution,
as computed by QMC for $\Delta \tau  D/\sqrt{2} = 1, 0.5$ and 
${\cal  G}_0^{met}(\tau,n_s=7)$ (from below). Parameters of the 
reference point are $\beta  D/\sqrt{2}=100, U=3.55  D/\sqrt{2}$. 
}
\end{figure} 

The situation becomes much more illuminating as we consider the
insulating solution ${\cal G}_0^{ins}(\tau)$. In this case, I have
computed $G^{\Delta \tau}(\tau)$ for $\Delta \tau=1,0.5, 0.25$.
The results for ${\cal  G}_0^{\mbox{new}}(\tau)$, again obtained
by inverse Fourier transformation, are shown
in Fig. 2:  it is evident that very large finite-$\Delta \tau$
effects are present.  Notice that {\em all} the curves ${\cal
G}_0^{\mbox{new}}(\tau,\Delta \tau)$ expose a plateau ${\cal G}_0
\sim constant $ for $\tau$ away from $0$ or $\beta$. Such a plateau
is characteristic of an insulating solution (cf. Fig. 1), and its
value {\em decreases} with the gap of the single-particle density
of states.  We thus arrive at the crucial observation that the
finite-$\Delta \tau$ effects bias the self-consistency condition
of the QMC algorithm towards the metallic solution.  Nevertheless,
we find back the ED solution as $\Delta \tau \rightarrow 0$.

\begin{figure}
\centerline{ \psfig{figure=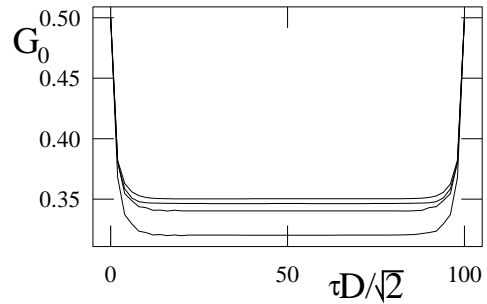,height=4cm} }
\caption{ 
${\cal  G}_0^{\mbox{new}}(\tau,\Delta \tau)$ of the insulating solution,
as computed by QMC for $\Delta \tau  D/\sqrt{2} = 1, 0.5, 0.25$ and 
${\cal  G}_0^{iso}(\tau,n_s=7)$ (from below) at the
reference point. 
}
\end{figure} 

In terms of the free-energy landscape mentioned earlier, this simply
means that the picture at finite $\Delta \tau$ is {\em tilted}. The 
QMC iteration is dragged  away from the insulating solution 
${\cal G}_0^{ins}$ into the metal.

Besides this effect, there seems to be no difference between
the iteration loop of ED and the one of QMC. Within a
deep basin of attraction of the insulating solution, a small drag
due to finite $\Delta \tau$ effects should only lead to a shift of the 
solution, and its 
stability should be preserved. This is exactly what
I have observed. I have run {\em full} iteration loops starting from 
${\cal G}_0^{ins}(\tau)$ for $\Delta \tau  D/\sqrt{2} = 1$ and $\Delta
\tau  D/\sqrt{2} = 0.5$. In the first case, the simulation moved away from the
insulation solution: After about $10$ iterations, the metallic
solution was approximately recovered. In contrast, 
for $\Delta \tau  D/\sqrt{2}= 0.5$, the insulating solution is
very clearly stable. A $40$-day simulation of this single
problem on a work station has obtained a very well-converged
self-consistent solution for ${\cal G}_0(\tau,\Delta \tau )$ within QMC. 
This solution makes no more reference to ED, but 
it of course resembles the curves shown in Fig. 2; the plateau
value is ${\cal G}_0(\tau  D/\sqrt{2}=50) \sim 0.30 $, 
comparable to the one-shot 
solution at the same value of $\Delta \tau$.

In my opinion, the present discussion of the discretization errors
and both the one-step iteration and the explicit self-consistent
Monte Carlo solution leave little room but to accept the  coexistence
at finite temperature.  It is evident that the Monte Carlo simulation
at the insulating solution has important finite-$\Delta \tau$
effects. These errors modify the qualitative aspects of the solution
only close to the phase boundaries where the basins of attraction
of the insulating solution are shallow, and small.  Therefore, the
discretization errors will be more pronounced close to $U_{c_1}$,
and also at higher temperature. This is what happened in the
simulation of ref. \cite{schlipf}, and what  led the authors to
wrong conclusions.

Acknowledgement: it is a pleasure to thank G. Kotliar and A. Georges for
helpful discussions.

\end{multicols}

\begin{thebibliography}{10}
\bibitem{footweb} The solutions ${\cal G}_0(\tau)$ are made 
available on request.
\end{thebibliography}
\end{document}